\documentclass[11pt]{iopart}
\usepackage{graphicx}
\usepackage[english]{babel}
\usepackage{amsfonts}
\usepackage{amssymb}
\expandafter\let\csname equation*\endcsname\relax
\expandafter\let\csname endequation*\endcsname\relax
\usepackage{amsmath}
\usepackage[ansinew]{inputenc}
\usepackage{multirow}
\usepackage{mathtools}
\usepackage{wasysym}
\usepackage{color}
\usepackage{verbatim}
\usepackage{bm}

\begin{document}

\title{Classical percolation fingerprints in the high temperature regime of 
quantum Hall effect}

\author{M. Fl\"{o}ser$^1$, B. A. Piot$^2$, C. L. Campbell$^2$, D. K. Maude$^2$,
M. Henini$^3$, R. Airey$^4$, Z. R. Wasilewski$^5$, S. Florens$^1$, T.
Champel$^6$}
\address{$^1$
Institut N\'{e}el, CNRS and Universit\'{e} Joseph Fourier, B.P.
166, 25 Avenue des Martyrs, F-38042 Grenoble, France}
\address{$^2$
Laboratoire National des Champs Magn\'{e}tiques Intenses, CNRS, UJF-UPS-INSA,
F-38042 Grenoble, France}
\address{$^3$
School of Physics and Astronomy, University of Nottingham, Nottingham NG7 2RD,
United Kingdom}
\address{$^4$
Department of Electronic and Electrical Engineering, University of Sheffield,
Sheffield S1 4DU, United Kingdom}
\address{$^5$Department of Electrical and Computer Engineering University of
Waterloo, Waterloo, ON, Canada}
\address{$^6$
Laboratoire de Physique et Mod\'elisation des Milieux Condens\'es,
CNRS and Universit\'{e} Joseph Fourier, B.P. 166, F-38042 Grenoble, France}

\date{\today }

\begin{abstract}
We have performed magnetotransport experiments in the high-temperature regime (up to 50 K) of
the integer quantum Hall effect for two-dimensional electron gases in
semiconducting heterostructures.
While the magnetic field dependence of the classical Hall law
presents no anomaly at high temperatures, we find a breakdown of the
Drude-Lorentz law for the longitudinal conductance beyond a crossover
magnetic field $B_c\simeq 1$ T, which turns out to be
correlated with the onset of the integer quantum Hall effect at low
temperatures. We show that the high magnetic field regime at $B>B_c$ can
be understood in terms of classical percolative transport in a smooth disordered potential.
From the temperature dependence of the peak longitudinal conductance, we extract
scaling exponents which are in good agreement with the theoretically expected
values. We also prove that inelastic scattering on phonons is responsible for
dissipation in a wide temperature range going from 1 to 50 K at high
magnetic fields.

\end{abstract}

\maketitle

\section{Introduction}

Two-dimensional electron gases (2DEGs) under perpendicular magnetic fields have
revealed at low temperatures a wealth of surprising transport properties \cite{vonK1980,Tsui1982,Ando1982}, which
are direct manifestations of quantum mechanics at the macroscopic scale.
Remarkably, in the same 2DEG system one may observe gradually by increasing the
magnitude of the magnetic field $B$ different quantum phenomena \cite{Prange1987}: Shubnikov-de
Haas (SdH) oscillations, followed by integer and then fractional quantum Hall effects (QHE).
All these effects capitalize on the quantization of the cyclotron orbital motion
resulting from the Lorentz force, which gives rise to discrete kinetic
energy levels $E_n=(n+1/2)\hbar \omega_c$ (with $n$ a positive integer,
$\omega_c=|e|B/m^{\ast}$ the cyclotron frequency, $e=-|e|$ the electron charge,
$m^{\ast}$ the effective mass, and $\hbar$ Planck's constant divided by $2
\pi$).
The QHE characterized by a spectacularly robust quantization of the Hall
conductance
in integral \cite{vonK1980} or fractional \cite{Tsui1982} multiples of $e^2/h$ differ
from the SdH (diffusive) regime by the quasi-absence of dissipation in the bulk, as
vindicated by the spectacular drop in magnitude of the longitudinal conductance
minima.
This transition is usually understood with the onset of a quasi-ballistic
transport regime \cite{But1988}, associated with the localization of all the
bulk states except at the center of a Landau level \cite{Prange1987} where a
diffusive electronic propagation throughout the system can set in only via
percolation. A semiclassical localization mechanism
\cite{Iordansky1982,Kazarinov1982,Trugman1983}
resulting from the decoupling of the (quantized) cyclotron motion with the
guiding center, which leads to a quasi-regular drift of the electronic states
along constant energy contours of the smooth disorder potential landscape, has
recently been confirmed \cite{Hash2008,Hash2012} by scanning tunneling spectroscopy in the integer QHE
regime. Percolative spatial structures for the local density of states taking
place at the transition between Hall plateaus have also been clearly identified
in this local probe experiment \cite{Hash2008}.
Signatures of percolation in transport properties have been mainly discussed in
the literature \cite{Wei1988,Wei1992,Li2005,Zhao2008,Li2009,Li2010,Saeed2011} at very low
temperatures (typically below $T=1$ K), when several quantum mechanical effects
(tunneling, quantum coherence, etc...) play a role
\cite{Chalker1988,Huckestein1995,Gammel1998,Cain2004,Kramer2005,Evers2008,Slevin2009,Amado2011,Obuse2012}
and complicate the analysis both theoretically and experimentally. However, if the localization
mechanism for the QHE is classical in nature, we expect these percolative features at high magnetic
fields to be also observable at much higher temperatures, in a classical transport
regime.

In this paper, we study the nature of the high magnetic field, high
temperature transport regime, combining experimental measurements in the 1-50 K
range with recent theoretical predictions \cite{Floser2011,Floser2012}.
We first identify a crossover magnetic field $B_c\simeq 1$ T above which
chaotic classical (diffusive) dynamics breaks down, that we correlate to the onset
of QHE at low temperature. This observation points to the common origin of
long-range disorder in suppressing the diffusive regime, both in the classical
and quantum realms.
For $B>B_c$ we observe various scaling laws that demonstrate the combined
role of phonon scattering and classical percolation in the transport properties.


\section{Observation of a breakdown of Drude-Lorentz law in high magnetic field}


The 2DEGs used in our study are delta-doped
Al$_{x}$Ga$_{1-x}$As/GaAs heterostructures patterned into a Hall
bar. The transport measurements were performed with a standard low
frequency lock-in technique for temperatures $T$ between 1.2 K and
50 K in a variable temperature insert, under magnetic fields up to
11 T. The first (second) sample has a mobility $\mu_e =3.3$
$\cdot$ 10$^5$ cm$^{2}$/Vs  at 1.2 K (8 $\cdot$ 10$^4$
cm$^{2}$/Vs), and an electron density $n_{e}=4$ $\cdot$ 10$^{11}$
cm$^{-2}$ (7 $\cdot$ 10$^{11}$ cm$^{-2}$). The two samples differ
also by their growth process since sample 1 is an heterojunction,
while sample 2 is a quantum well~\cite{Sample2}. A particular
attention has been paid to the mobility range of the samples chosen
to investigate the high-temperature regime of the QHE. On the one
hand, high enough mobility was required by the need to clearly
separate the crossover magnetic field $B_c$ for which the
classical localization is expected to set in and the $B$-scale
where the quantization of the cyclotron motion starts to be felt
(typically, $B\simeq0.2$ T corresponds to $\hbar \omega_c \simeq 4
k_B T$ for the lowest temperatures studied here). On the other
hand, very high mobility samples had to be avoided because of the
importance of many-body effects (such as e.g. spin splitting)
blurring the physics under consideration.


\begin{figure}[t]
\includegraphics[width=0.7\linewidth]{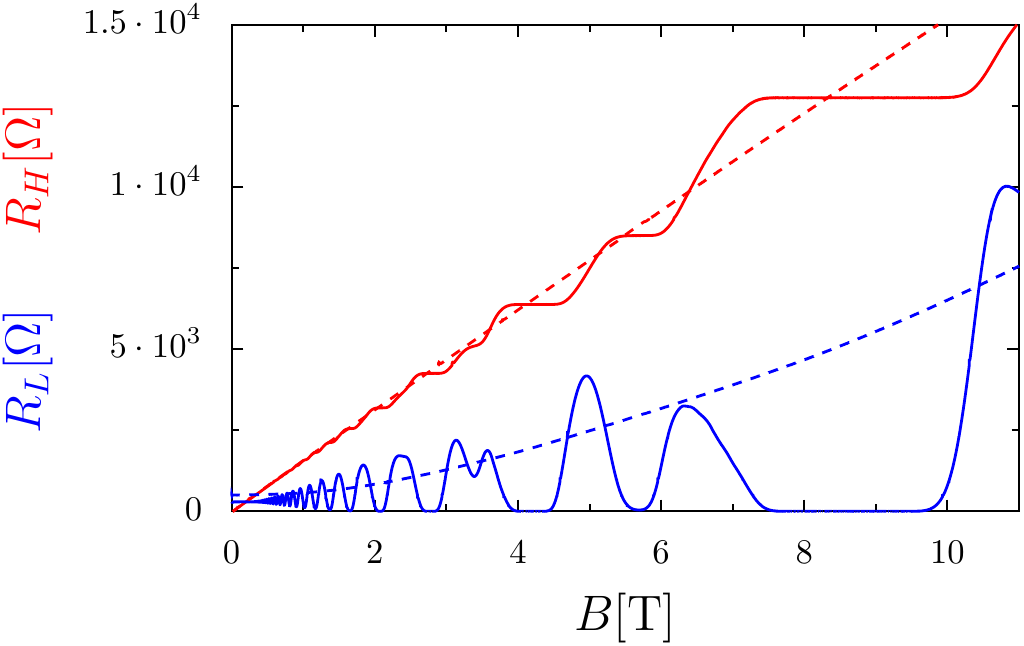}
\caption{
Longitudinal $R_L$ (bottom curves) and Hall $R_H$ (top
curves) resistances as a function of magnetic field at $T=1.2$ K (solid lines)
and $T=47$ K (dashed lines) for sample 1.}
\label{Fig1}
\end{figure}

The $B$-dependences of the Hall and longitudinal resistances for
sample 1 are shown in Fig. \ref{Fig1} at low and high temperatures
($T=1.2$ K and $T=47$ K, respectively). The low-$T$
field-dependence is quite standard with the appearance of
well-formed plateaus for Hall resistance $R_H$ which are
accompanied by strong oscillations of the longitudinal resistance
$R_L$ with vanishing minima for fields $B \gtrsim 1 $ T. We note
that peaks of $R_L$ start to be spin-resolved for $B \geq 3$ T at
this low temperature due to the critical many-body enhancement of
the spin gap.
At high $T$, $R_H$
becomes structureless and exhibits a linear dependence in field as expected from
classical Hall's law.
The observed longitudinal magnetoresistance shows a richer and
more instructive field dependence.  While $R_L$ saturates
according to the classical behavior at low magnetic fields, it
shows for $B \gtrsim 1 $ T a steady super-linear increase with the
magnetic field. A similar positive and slightly non-linear
magnetoresistance is found for sample 2, see Appendix.

\begin{figure}[t]
\includegraphics[width=0.5\linewidth]{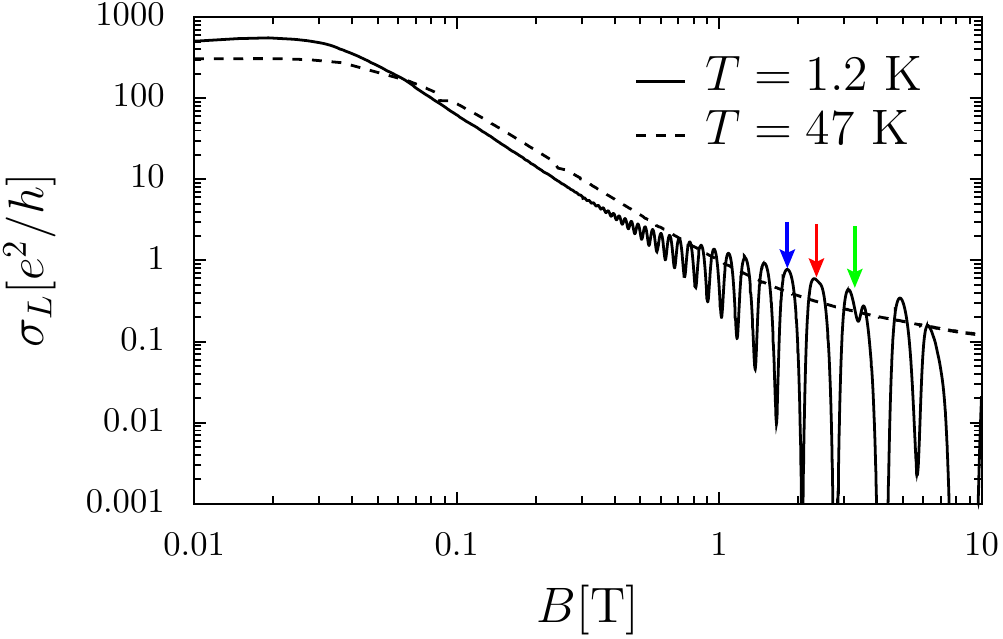}
\includegraphics[width=0.5\linewidth]{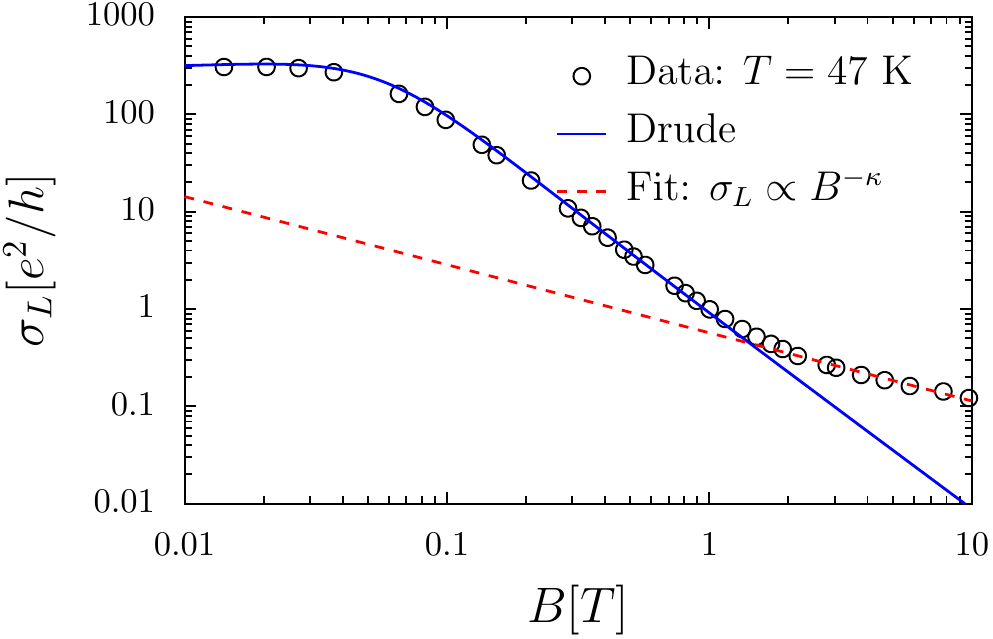}
\caption{
Left panel: Longitudinal magnetoconductance $\sigma_L$ for sample 1 as a function of magnetic field at
$T=1.2$ K (solid line) and $T=47$ K (dashed line), correlating the breakdown
of mild SdH oscillations in the quantum regime to the one of Drude-Lorentz law in the
classical limit. Arrows denote a set of conductance peaks associated to the
quantum Hall transitions examined in Fig.~\ref{Fig3}.
Right panel: study of the high temperature data (circles).
Drude's law (\ref{Drude}) is well obeyed for $B<B_c=1$ T (top solid line),
while an anomalous power law $B^{-\kappa}$ with $\kappa=0.7 \pm 0.1$ is seen
at $B>B_c$ (bottom dashed line).}
\label{Fig2}
\end{figure}

This transition regime for the longitudinal dissipative transport coefficient is
better analyzed in terms of conductance $\sigma_L$ rather than resistance, see
Fig.~\ref{Fig2}.
The magnetic field
separating the SdH regime from the integer QHE regime
is identified at low $T$ in the range 1-2 T by the exponential drop of
$\sigma_L$ in its minima values.
It is interesting to correlate this observation with the measurement
performed at high $T$, especially on a logarithmic scale as shown in
Fig.~\ref{Fig2}.
Classically, $\sigma_L$ is expected to obey the Drude-Lorentz law,
\begin{eqnarray}
\sigma_L=\frac{n_e e^2}{m^{\ast}} \, \frac{\tau_{\mathrm{tr}}}{1+(\omega_c
\tau_{\mathrm{tr}})^2},
\label{Drude}
\end{eqnarray}
where  $\tau_{\mathrm{tr}}$ is the transport time determined by the combined
scattering on the (random) impurity potential and by phonons. As long as
$\omega_c \tau_{\mathrm{tr}} \ll 1$, $\sigma_L$ remains constant, while a quadratic
decrease $\sigma_L \propto B^{-2}$ is
expected when $\omega_c \tau_{\mathrm{tr}} \geq 1$. Drude's law
(\ref{Drude}), also associated to a constant magnetoresistance $R_L$ in Fig.~\ref{Fig1}, is
well verified up to the magnetic field
$B_c \sim 1$ T, with $\tau_{\mathrm{tr}}=7.4$ $\cdot$ 10$^{-12}$ s at $T=47$ K.
Above $B_c$, an anomalous power-law dependence,
namely $\sigma_L \propto B^{-\kappa}$ with $\kappa \approx 0.7 \pm 0.1$, is revealed by
the logarithmic plot of Fig.~\ref{Fig2}.  Interestingly, we note that this
breakdown of Drude's law at high $T$ appears correlated to the onset of the
QHE at low $T$. Because we are working at a relatively high $T$,
it is very likely that this breakdown has a purely classical origin.

A linear classical magnetoresistance in several high-mobility samples at
high $T$, qualitatively similar to that shown in Fig. \ref{Fig1}, has already
been reported several years ago by R\"otger {\em et al.}~\cite{Roetger}.
These observations were correlated to the empirical proportionality relation
between longitudinal and Hall resistances,
$R_L\propto B\, \mathrm{d} R_H/\mathrm{d} B$, originally pointed out in the
low temperature quantum Hall regime in Ref.~\cite{Chang}. Indeed, in
the classical Hall regime where $R_H \propto B$, this empirical relation
predicts a purely linear longitudinal magnetoresistance. It has been argued that
this linear behavior at high temperature could be explained by macroscopic density inhomogeneities
in the sample~\cite{Hirai}. Nevertheless, we note that this interpretation leaves unexplained
the clear transition to the constant Drude-Lorentz resistance at low magnetic field
as observed in our data, see Fig.~\ref{Fig1}.

More recent studies by Renard {\em et al.}~\cite{Renard2004}
evidenced obvious non-linearities in the high $T$
magnetoconductance, with a dependence $R_L \propto B^{\alpha}$ and
an exponent $\alpha\simeq 0.9-1.1$. This translates into a
high-field dependence for the conductance $\sigma_L \propto
B^{-\kappa}$, where the exponent $\kappa \approx 0.9$. This turns
out to be higher than the value of $0.7\pm 0.1$ that we have
extracted for sample 1, but all these measurements concur to
invalidate both the high-field Drude-Lorentz law $\sigma_L\propto
B^{-2}$ and the empirical relation (between $R_H$ and $R_L$)
leading to $\sigma_L\propto B^{-1}$. The appearance of a
non-integer exponent $\kappa$ is rooted in the percolative nature
of transport in actual samples, which is the main issue addressed
in this paper.

Note that in our other sample 2, we obtain $\kappa \approx 0.8\pm0.1$ (see
Appendix), i.e., again a non-linearity of the magnetoconductance. The error
bars on the high-field scaling exponent $\kappa$ are mainly due to the limited
range of magnetic field which can be used, making a very accurate
extraction of $\kappa$ difficult from a simple field-dependence.
Indeed, even at $T=50$ K, Landau level quantization gives rise to noticeable
quantum oscillations of $\sigma_L$ superposed to the power-law scaling
background $\sigma_L\propto B^{-\kappa}$, since $\hbar \omega_c \simeq 4 k_B T$
for magnetic field $B>8$ T.

To understand these discrepancies between the different extracted exponents, it
is necessary to better characterize the disorder potential in the different
samples. The analysis of the magnetic field dependence of SdH oscillations at
temperature $T=1.2$ K leads  to a quantum lifetime $\tau_q=1.2$ ps for sample 1,
and $\tau_q=0.6$  ps  for sample 2. The transport lifetimes calculated at the
same temperature from mobility measurements  ($\tau_{\mathrm{tr}}=12.7$ ps in
sample 1, and $\tau_{\mathrm{tr}}=3$ ps in sample 2) are in both samples larger
than the quantum lifetimes, what shows the dominant contribution of long range
scattering due to a long range disorder potential. The ratio of the transport
time to the single-particle scattering time ($\tau_{\mathrm{tr}}/\tau_q=10.6$
in sample 1, $\tau_{\mathrm{tr}}/\tau_q=5$ in sample 2) is
however not as strong for sample 2, indicating the non negligible contribution
of short range scattering mechanisms in the sample (such as, e.g., surface
roughness), which could be at the origin of the observed discrepancies. With the
help of a theoretical analysis of the data to be developped in the next section,
we will first confirm the above exponent values by using in addition the temperature
scaling of the magnetoconductance, which will further support the present
interpretation for the exponent dispersions.

\section{Theoretical analysis of magnetotransport data}

A breakdown of law (\ref{Drude}) has been predicted in a few theoretical papers \cite{Fog1997,Polyakov2001}
addressing long-range disorder at large magnetic field.
Essentially, Drude-Lorentz formula relies on classical diffusive transport
with chaotic electronic motion due to elastic collisions on impurities at low magnetic fields. When the
cyclotron radius becomes basically smaller than the correlation length of the disorder potential, this evolves at high
magnetic fields into a quasi-ballistic transport regime with a regular motion
of the guiding center along the constant energy contours of the disorder potential
landscape, which follows mainly closed trajectories. Macroscopic transport
then only takes place by following an extended percolating backbone occurring at a
single critical energy and passing through many saddle points of the disorder
landscape. The fractal nature of the percolating contour is expected to give rise
to non-trivial universal exponents in the temperature and magnetic field dependences
of $\sigma_L$, as reported here at high magnetic fields.
However, it is worth stressing that the percolating contour alone is not
sufficient to allow macroscopic transport, since the guiding center drift velocity
vanishes at the saddle-points of the disorder landscape. Different microscopic
dissipative processes may a priori be at play to provide a finite drift velocity at
these transport bottlenecks, an issue that will be clarified in the present
work.

It has been argued in Ref. \cite{Fog1997}, that the high temperature
crossover field $B_c$ for the Drude-Lorentz breakdown should be quite close to
the low temperature transition between the SdH and QHE regimes. Our experiments
in the two samples corroborate this scenario. However, the classical prediction \cite{Fog1997} of
an exponential suppression of $\sigma_L$ with $B$ above $B_c$, which is based on a
mechanism of dissipative transport via a stochastic layer around the percolating contour
resulting from elastic scattering on the disorder random potential only, is not consistent
with the power-law decrease seen in Fig. \ref{Fig2}, hinting at more efficient
relaxation processes.


We now provide detailed theoretical analysis of our experimental data.
The high-$B$ percolative transport regime  can be described in terms of a Ohm's law
involving a local conductivity tensor
\cite{Ruzin1993,Simon1994,Floser2011,Floser2012}, which takes the form
\begin{eqnarray}
\hat{\sigma}({\bf r})=\left(
\begin{array}{cc}
\sigma_0 & - \sigma_H({\bf r})\\
\sigma_H({\bf r})
&
\sigma_0
\end{array}
 \right),
\label{tensor}
\end{eqnarray}
where $\sigma_0$ encodes dissipative processes (assumed to be uniform), and
$\sigma_H({\bf r})$ is the local Hall component, whose spatial dependence
originates from charge density fluctuations due to disorder $V({\bf r})$ in the
sample. The local conductivity model expresses  the inhomogeneous nature of the
high-magnetic field transport, which  results from the formation of local
equilibrium, and is valid at temperatures high enough so that phase-breaking
processes, such as electron-phonon scattering, occur on length scales that are
shorter than the typical variations of disorder. The Ohmic conductivity
$\sigma_{0}$ is assumed very weak [i.e., $\sigma_{0} \ll  \sigma_{H} ({\bf r})$]
but finite. A priori, it may be  due to other scattering mechanisms than elastic
impurity scattering such as
electron-phonon scattering \cite{Ruzin1993,Polyakov1995,Floser2011,Floser2012}.  It
has been found \cite{Simon1994,Floser2011} from model (\ref{tensor}) that the
longitudinal conductance scales as
\begin{eqnarray}
\sigma_L =C \left[\sigma_0 (T,B)\right]^{1- \kappa} \left| \frac{e^2}{h}
\sqrt{\langle V^2 \rangle} \sum_{n=0}^{\infty} n'_F(E_n-\mu)
\right|^{\kappa}\!\!\!,
\label{sigmaL}
\end{eqnarray}
where $\kappa$ is a non-trivial exponent previously conjectured \cite{Simon1994}
to be $\kappa=10/13 \approx 0.77$, a result confirmed recently by a diagrammatic
approach \cite{Floser2011}. Here $C$ is a nonuniversal dimensionless constant,
$\mu$ is the chemical potential, and $n'_F$ is the derivative of the Fermi-Dirac
distribution function. Formula (\ref{sigmaL}) has been established under the
assumption that $\sigma_H({\bf r})$ follows {\em linearly} the spatial fluctuations of
disorder $V({\bf r})$, what requires a sufficiently high $T$. Typically, it does not
describe the low temperature regime~\cite{Dykhne1994,Floser2012} when the peak conductance
starts to level off around conductance values of the order of $e^2/2h$ (per spin).
Another important limitation of formula~(\ref{sigmaL}) is the assumption of a Gaussian correlated
disorder with a single correlation length, which is valid for impurities located far away
from the gas and in absence of sources of short-range impurity scattering.
For instance, sample 2 (a quantum well) is likely characterized 
by multi-scale disorder, which would involve extra (unknown) microscopic
parameters in the modelling , making difficult a quantitative comparison to theory
(see discussion in Appendix).

In addition, model~(\ref{tensor}) leads~\cite{Simon1994,Floser2012} to the
conventional classical Hall's law at $T\gg\hbar\omega_c$, where fingerprints of percolations are absent. It
predicts therefore a slight breakdown of the empirical proportionality law
(between $R_L$ and $R_H$) due to the non-linearity related to the
non-integer exponent $\kappa$. Indeed, this gives at high magnetic field
$R_L \propto \sigma_L R_H^2 \propto B^{2-\kappa}$, which is non-linear for
$\kappa\neq1$. It has been argued~\cite{Simon1994} however that the presence of
disorder on multiple length scales tends to increase the exponent $\kappa$
towards 1, thus recovering the empirical proportionality law in the case of more
inhomogeneous samples.

Expression (\ref{sigmaL}) yields oscillations of $\sigma_L$ in $B$ in the
percolation regime  when $k_B T \leq \hbar \omega_c$, which are superposed on a
high-$T$ classical background conductance (obtained for $k_B T > \hbar
\omega_c$)
\begin{eqnarray}
\sigma_{\text{bg}}(T,B) =C \left[ \sigma_0(T,B)\right]^{1-\kappa}
\left[ \frac{e^2}{h} \frac{\sqrt{\langle V^2 \rangle }}{ \hbar \omega_c} \right]^{\kappa}.
\label{sigmab}
\end{eqnarray}
It has been predicted \cite{Floser2011} that the peak values for $\sigma_L$  are
given by the formula
\begin{eqnarray}
\sigma_{L}^{\mathrm{peak}}= \sigma_{\text{bg}}(T,B) \!\!
\left[ 1+\sum_{l=1}^{\infty} \frac{4 \pi^2 l k_B T}{\hbar \omega_c} \text{csch}
\left( \frac{2 \pi^2 l k_B T}{\hbar \omega_c}\right) \right]^{\kappa}.
\label{sigmaLpeak}
\end{eqnarray}
We deduce from Eq. (\ref{sigmab}) that in the classical percolative regime
$\sigma_L$ should scale in $B$ as $\sigma_L \propto \left[
\sigma_0(T,B)\right]^{1-\kappa} \, B^{-\kappa}$.
Provided that $\sigma_0$ is quasi-constant in magnetic field, we find a power-law
scaling $\sigma_L\propto B^{-\kappa}$, which globally agrees with the experimental
$B$-dependences reported above.

\begin{figure}[tb]
\includegraphics[width=0.7\linewidth]{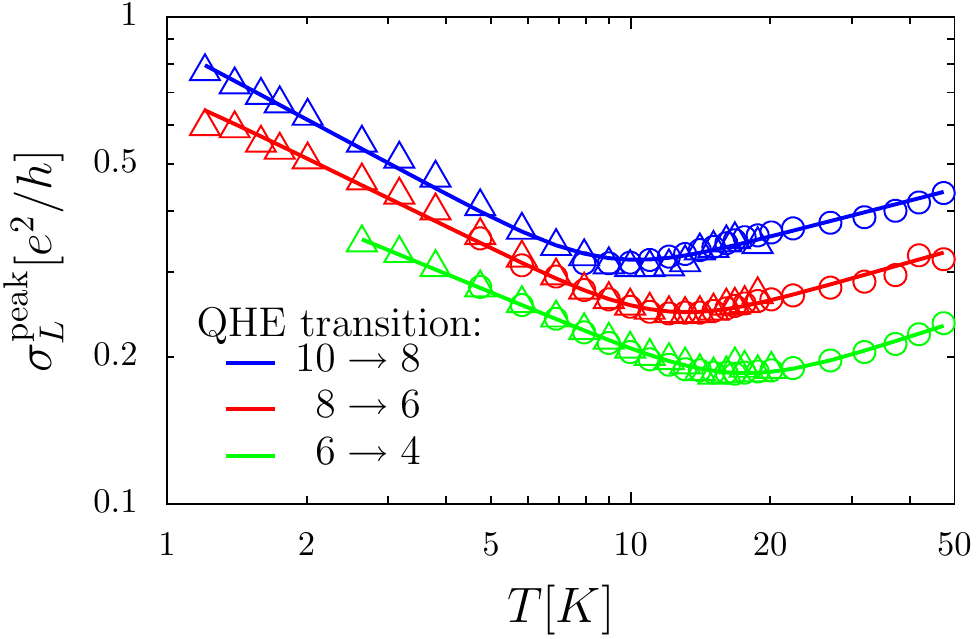}
\caption{
Temperature dependence of the peak longitudinal conductance at the QHE
transitions with filling factors 10 $\rightarrow$ 8,  8 $\rightarrow$ 6, and 6 $\rightarrow$ 4 (top
to bottom) for sample 1, as indicated by arrows on Fig.~\ref{Fig2}.
Triangles design values measured at the conductance peaks, and circle values
taken at fixed $B$ field (see text). The lines are the fit curves with Eq.
(\ref{sigmaLpeak}) and the fit parameters given in Table \ref{Tab2}.
}
\label{Fig3}
\end{figure}

We now consider the temperature dependence of the conductance in the
percolative transport regime, focusing on the spin-unresolved conductance
peaks indicated by arrows on Fig.~\ref{Fig2}. In the low temperature
range, we monitor and follow the conductance maxima, which are plotted
as triangles on Fig.~\ref{Fig3}.
At higher $T$, these peaks are washed out and cannot be followed individually
anymore, but the magnetic field may be then kept constant, as the conductance becomes
weakly field-dependent (circles on Fig.~\ref{Fig3}).  We also note that the opening of the spin gap at the
highest magnetic fields considered here limits the temperature range where
Zeeman and many body effects can be neglected.
We first observe on Fig.~\ref{Fig3} the presence of a pronounced minimum
at a temperature that perfectly correlates for each peak with the scale $T^{\ast}=\hbar
\omega_c/(4 k_B)$ where quantized Landau levels start to emerge, a striking effect
that went previously unnoticed to our knowledge.
%
In addition, two different power-law scalings (with a negative power at
$T<T^\ast$ and positive one at $T>T^\ast$) are clearly seen, that we would
like to attribute to the critical state associated to classical percolation. It can be easily noted from Fig.~\ref{Fig3} that the classical conductance  seems to scale as $\sigma_L \propto T^{1-\kappa}$ for $T>T^\ast$, what implies a linear temperature dependence for $\sigma_0$ according to Eq. (\ref{sigmab}).

In order to understand this rich temperature dependence of the conductance peaks, we now need to characterize microscopically the missing piece in formulas~(\ref{sigmab})-(\ref{sigmaLpeak}),
namely the  dissipative contribution $\sigma_0$. Its origin may be the inelastic electrons-phonons scattering, as has been put forward in many theoretical papers
\cite{Polyakov1995,Fogler1995,Fogler1996}.
If we assume that the electrons undergo a large number of scattering events on
the phonons, i.e., their rate
$\tau_{\mathrm{ph}}^{-1}$ is much higher than the characteristic frequency of
drift motion, we can estimate the \textit{short-distance} dissipative conductivity
$\sigma_0$ using the Drude-Lorentz formula
\begin{eqnarray}
\sigma_0=\frac{n_e e^2 }{m^{\ast}} \frac{\tau_{\mathrm{ph}}}{1+(\omega_c \tau_{\mathrm{ph}})^2},
\label{sigma0}
\end{eqnarray}
where $\tau_{\mathrm{ph}}$ is the electron-phonon scattering time.  Our
estimation of $\tau_{\mathrm{ph}}$ in the regime of the quantum Hall effect
follows that from Ref. \cite{Zhao} using Fermi's golden rule \cite{note2}.
This yields \cite{Zhao} a
scattering rate $\tau_{\mathrm{ph}}^{-1} \propto B^2 T$. Inserting this result
in Eq. (\ref{sigma0}), we obtain that $\sigma_0$ indeed scales linearly in
temperature and is independent of magnetic field whenever $\omega_c
\tau_{\mathrm{ph}} \gg 1$, vindicating an asumption made earlier. It is worth
mentioning that the temperature range where the linear $T$-dependence of
$\tau_{\mathrm{ph}}^{-1}$ is valid increases at high magnetic fields
\cite{Zhao}. The saturation of $\tau_{\mathrm{ph}}^{-1}$ due the spontaneaous
phonon emission \cite{Levinson} is expected to take place typically at low
temperatures in the Kelvin range, which is beyond the present analysis.

From the established $\sigma_0\propto T$ law,
formulas~(\ref{sigmaL}-\ref{sigmaLpeak})
predict that the cyclotron energy separates two distinct physical regimes of
transport: i) at $T>T^\ast$ percolation of guiding centers carrying a classical cyclotron
motion occurs, leading to $\sigma_L^{\mathrm{peak}}\propto
T^{1-\kappa}=T^{3/13}$. This is easily established by noting that the r.h.s.
term under brackets in Eq.~(\ref{sigmaLpeak}) becomes constant in temperature;
ii) at $T<T^\ast$, the cyclotron motion becomes quantized while the transport of
the guiding center remains classical, changing the $T$-dependence into
a {\it negative} power-law $\sigma_L^{\mathrm{peak}} \propto T^{1-2\kappa}=T^{-7/13}$.
This new power-law derives directly from Eq.~(\ref{sigmaL}), noting that the
Fermi function derivative behaves as $1/T$ at low temperature.
The assumption of a dissipation mechanism with phonons combined with formula~(\ref{sigmaLpeak}) 
thus not only describe qualitatively our data for the temperature dependence of
$\sigma_L^{\mathrm{peak}}$,
but also provides a very precise way of extracting the classical exponent
$\kappa$.

The fits performed on Fig.~\ref{Fig3} for three successive plateau transitions
yield the results given in Table \ref{Tab2}. This comparison is performed by
considering $\sigma_{\mathrm{bg}}=A T^{1-\kappa}$, where the amplitude $A$ and
the critical exponent $\kappa$ are the only fitting parameters. We find that the
agreement between the experimental data and the fitting curve is excellent for
almost two decades in temperatures.  The consistency of the results obtained from both the temperature and magnetic field dependences of $\sigma_L$ also provides good confidence in the theory.
For the peak located at the transition between filling factors 10 to 8 ($B=1.7$ T),
the values extracted for the critical exponent $\kappa$ are very close to the
theoretical prediction $\kappa=10/13\simeq0.77$.
We note that the agreement weakens for lower filling factors, where spin splitting effects
may play an increasing role that we have not accounted for in the calculation.

%
%
%
%
\begin{table}[h]
\begin{tabular}{l c c}
\hline \hline
$B$ [T]  &  \hspace*{0.2cm} $A$ [in units of $h/e^2 \, $ K$^{\kappa-1}$] \hspace*{0.2cm} & $\kappa$  \\
\hline
1.7 & 0.16  & 0.76 $\pm$ 0.03 \\
2.3  & 0.115 & 0.73 $\pm$ 0.03 \\
 3.3 & 0.071 & 0.70 $\pm$ 0.02
\\
\hline \hline
\end{tabular}
\caption{Fit parameters for Fig. \ref{Fig3}.}
\label{Tab2}
\end{table}

Turning to sample 2, the analysis of the temperature dependence of
each studied conductance peak yields $\kappa \approx 0.85\pm0.05$ (for more details, see
Appendix), which is consistent with the exponent extracted from the
$B$-dependence. This slightly larger value for the classical percolation exponent
in sample 2 could be due to the presence of disorder fluctuations
over different length scales, which have the effect of increasing the effective
percolation exponent as discussed in Ref. \cite{Simon1994}. Indeed, sample
2 is a narrow (8.2 nm) quantum well, which, in addition to impurity potential, exposes
the electron gas to interface roughness.




\section{Summary}
In conclusion, we have shown that the onset of the quantum Hall effect at low
temperatures is directly linked with a breakdown of the classical diffusive
regime at high temperatures. We have pointed out that temperature and magnetic
field dependences of the longitudinal conductance follow peculiar scaling
power-laws in the breakdown regime. We have found a good agreement between
theory and experimental data, which confirms that transport is dominated by
classical percolation in a wide temperature range going from 1 to 50 K in 2DEGs
at high magnetic fields.
The analysis also shows that the interaction of phonons with bulk drift states provides
the dominant dissipation mechanism at play in this temperature regime.


\section*{Acknowledgments}
We thank D. Basko and V. Renard for interesting discussions, and ANR ``METROGRAPH''
under Grant No. ANR-2011-NANO-004-06 for financial support.

\appendix
\section*{Appendix}

We present in this appendix the results obtained for sample 2
(quantum well), which has been grown differently from sample 1 (heterojunction).
We will follow the logic of the main text, by first considering the magnetic field
dependence, and finally the temperature behavior of the longitudinal conductance in
the high temperature regime of the quantum Hall effect.

\subsection{Magnetic field dependence of transport coefficients in sample 2}

\begin{figure}[htb]
\includegraphics[width=0.7\linewidth]{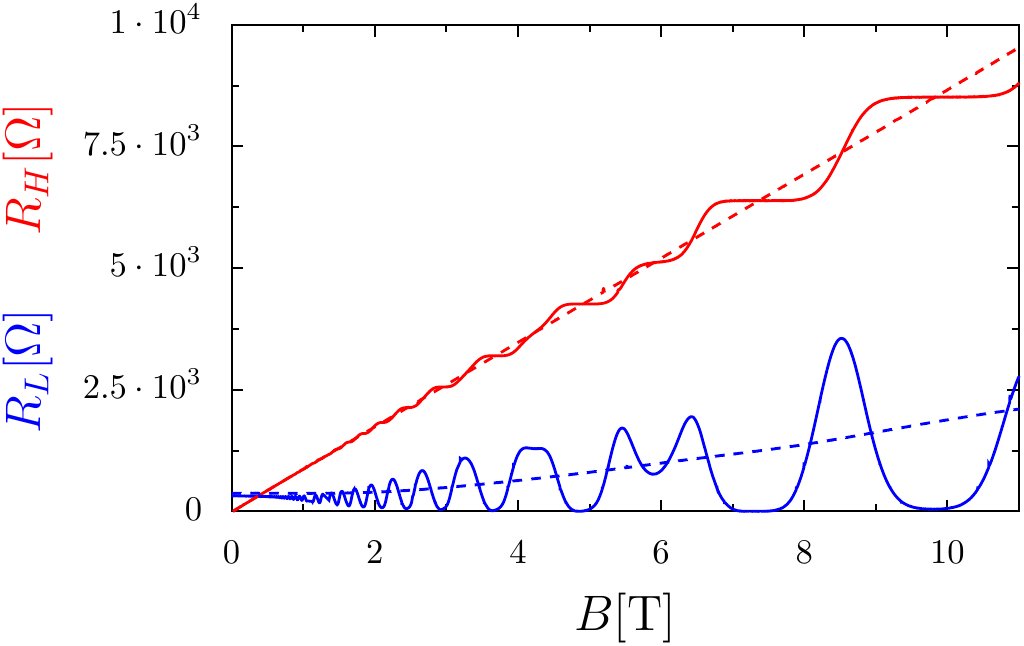}
\caption{
Longitudinal $R_L$ (bottom curves) and Hall $R_H$ (top
curves) resistances as a function of magnetic field at $T=1.2$ K (solid lines)
and $T=50$ K (dashed lines) for sample 2.}
\label{Fig1SupInfo}
\end{figure}

In Fig. \ref{Fig1SupInfo}, we present the magnetic field dependences of the Hall
and longitudinal resistances observed in sample 2, at temperatures $T=1.2$ K and
50 K. At low temperatures, the integer quantum Hall effect is clearly observed
above magnetic fields $B \gtrsim 2$ T. As reported in Fig.~\ref{Fig1},
the longitudinal resistance of the second sample also displays a steady increase
with the magnetic field for fields $B \gtrsim 2$ T at high temperatures, a
behavior which turns out to be again correlated with the onset of the quantum
Hall effect at low temperatures. We note that the transition from the
Shubnikov-de Haas regime to the quantum Hall regime occurs at a slightly higher
crossover field than in sample 1. This observation is consistent with the fact
that the mobility in sample 2 is smaller than in sample 1 (which is likely
characterized by a smoother disorder potential landscape).

\begin{figure}[htb]
\includegraphics[width=0.5\linewidth]{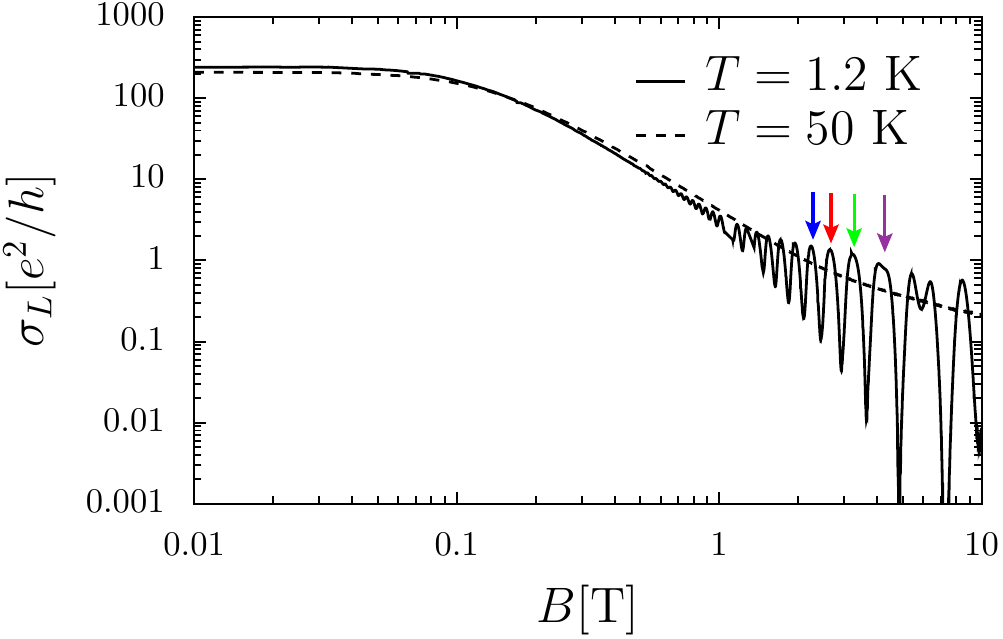}
\includegraphics[width=0.5\linewidth]{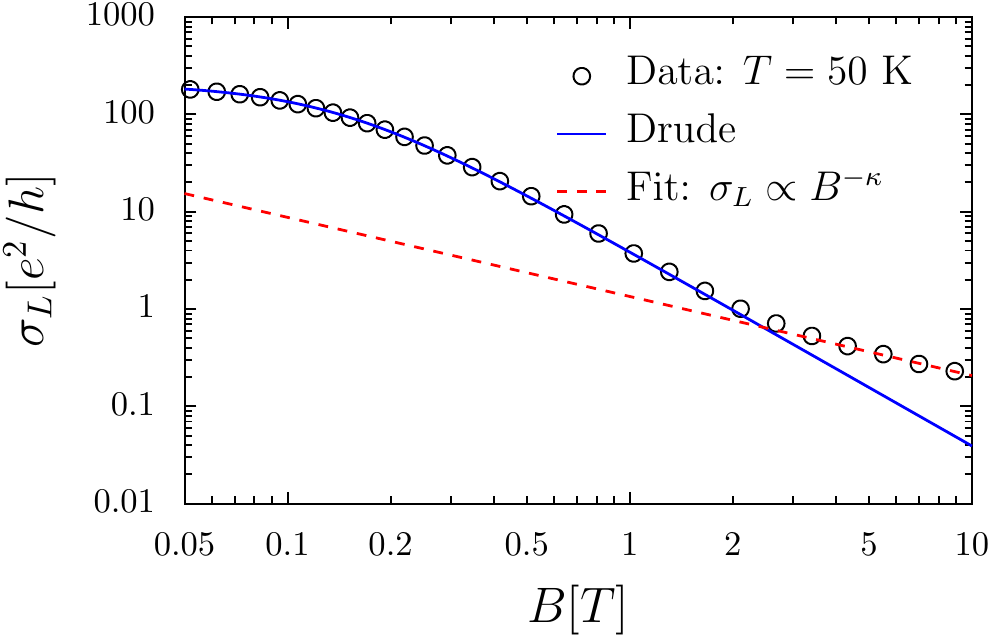}
\caption{
Left panel: Longitudinal magnetoconductance $\sigma_L$ for sample 2 as a function of magnetic field at
$T=1.2$ K (solid line) and $T=50$ K (dashed line), correlating the breakdown
of mild SdH oscillations in the quantum regime to the one of Drude's law in the
classical limit. Arrows denote a set of conductance peaks associated to the
quantum Hall transitions examined in Fig.~\ref{Fig3SupInfo}.
Right panel: study of the high temperature data (circles).
Drude law Eq.~(\ref{Drude}) is well obeyed for $B<B_c=2$ T (top solid line),
while an anomalous power law $B^{-\kappa}$ with $\kappa=0.8\pm0.1$ is seen
at $B>B_c$ (bottom solid line) for the studied sample.}
\label{Fig2SupInfo}
\end{figure}

The $B$-dependence of the longitudinal transport coefficient can be better
analyzed quantitatively in terms of conductance in a logarithmic scale, as shown
in Fig. \ref{Fig2SupInfo}. The Drude-Lorentz law Eq. (1) perfectly
describes the low-field part of the conductance and  yields
$\tau_{\mathrm{tr}}=2.85 \cdot 10^{-12}$ s at $T=50$ K. At higher magnetic fields  $B
\gtrsim 2$ T, the Drude-lorentz law also clearly breaks down in sample 2, with
the scaling dependence $\sigma_L \propto B^{-\kappa}$ and $\kappa \approx 0.8$.

\subsection{Temperature dependence of the peak longitudinal conductance in sample 2}

As discussed in the main text, a more accurate determination of the percolation
exponent is most conveniently obtained from the temperature dependence of the
longitudinal conductance at peak values. At temperatures of the order of the
cyclotron gap ($k_B T \gtrsim \hbar \omega_c/4$), the maxima are washed out, so
that the temperature dependence is then followed by working at constant
$B$-field. The conductance peaks are studied in the breakdown regime at fields
$B > B_c$, and are chosen such that the (unknown) temperature dependence of the
spin gap does not play a role. Obviously, this puts a stringent constraint on
the allowed peaks. The arrows in Fig. \ref{Fig2SupInfo} indicate the peaks that
we have analyzed in detail for sample 2.

\begin{figure}[tb]
\includegraphics[width=0.7\linewidth]{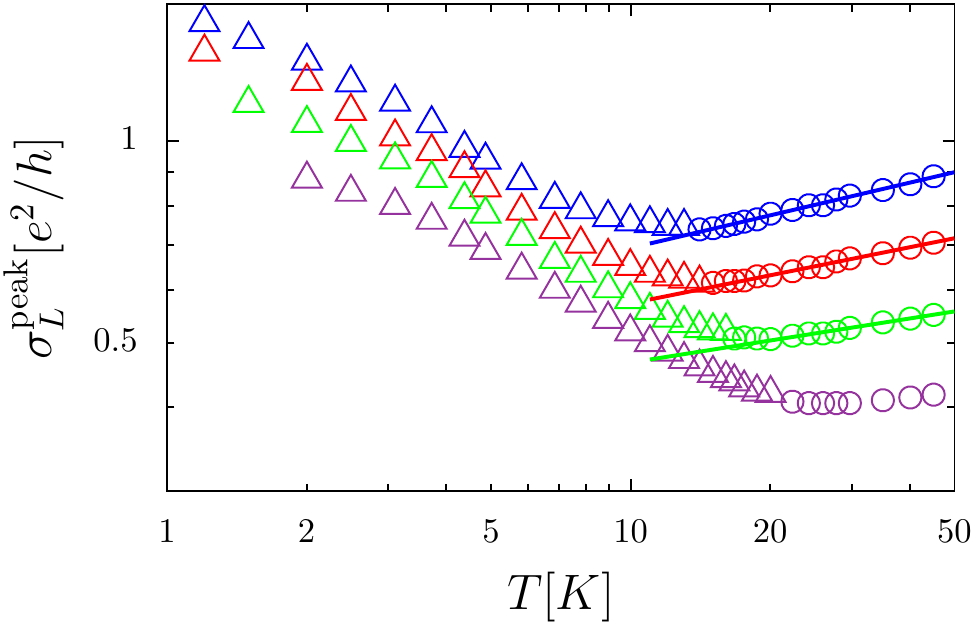}
\caption{
Temperature dependence of the peak longitudinal conductance for sample 2. The
selected peaks are indicated by arrows on Fig.~\ref{Fig2SupInfo}.  Triangles
design values measured at the conductance peaks, and circle values taken at
fixed $B$ field. The lines are fit to $\sigma_L =A T^{1-\kappa}$, with
parameters given in Table \ref{Tab2SupInfo}. The lower curve (for $B=4.28$ T)
cannot be reliably fitted due to the reduced range in temperature.
}
\label{Fig3SupInfo}
\end{figure}

The temperature dependences of the selected peaks are shown in Fig.
\ref{Fig3SupInfo} where a double logarithmic scale is used to better display the
scaling laws. As reported for sample 1 in the main text, a minimum at the
characteristic temperature scale $T^{\ast}=\hbar \omega_c/(4 k_B)$ separating
two temperature regimes with different scalings is seen for each studied peak in
sample 2. This is again qualitatively consistent with
formula~(\ref{sigmaLpeak}).
Note however, that in contrast to sample 1, the low temperature conductance
peaks exceed the maximum value $e^2/h$ (in the spin degenerate case)
expected for a smooth potential~\cite{Dykhne1994,Floser2012}, see Fig.~\ref{Fig3SupInfo}.
Formula~(\ref{sigmaLpeak}) is thus not appropriate to fit quantitatively the whole
temperature range, as performed successfully with sample 1.
We remind that formula~(\ref{sigmaLpeak}) is derived for a smooth disorder potential
characterized by a single length scale, an
assumption which may not be totally correct for sample 2. Indeed the latter is a
narrow quantum well with important surface roughness (and correspondingly lower
mobility), which is not accounted for in our model.
We thus confine our study of the critical exponents to the high temperature
regime above $T^{\ast}$, where density fluctuations are thermally smeared, so
that equation~(\ref{sigmab}) should still be valid in this regime.
The resulting high-temperature longitudinal conductance is expressed as $\sigma_L =A T^{1-\kappa}$,
where $A$ is an amplitude dependent on magnetic field that will be taken as fit parameter.
A second fit parameter is given by the critical exponent $\kappa$, and the
results are compiled in Table~\ref{Tab2SupInfo}.

\begin{table}[htb]
\begin{tabular}{l c c}
\hline \hline
$B$ [T]  &  \hspace*{0.2cm} $A$ [$h/e^2 \, $ K$^{\kappa-1}$] \hspace*{0.2cm} & $\kappa$  \\
\hline
2.2 & 0.47 & 0.84 $\pm$ 0.05 \\
2.65 & 0.41 & 0.86 $\pm$ 0.05 \\
3.23 & 0.36 & 0.89 $\pm$ 0.06 \\
\hline \hline
\end{tabular}
\caption{Fit parameters for Fig. \ref{Fig3SupInfo}.}
\label{Tab2SupInfo}
\end{table}

As a result, the extracted values for the exponent $\kappa$ in sample 2 are still
close to the theoretical prediction $\kappa \approx 0.77$, although we note here a
slight overestimation.
The systematic bias in $\kappa$ observed in sample 2 could also be
attributed to the more pronounced roughness of the disordered potential
landscape compared to sample 1, as discussed above.


\begin{thebibliography}{99}
\bibitem{note2}
This derivation is close to the standard textbook version \cite{Gant} of
electron-phonon scattering, which has to be extended to take into account the
Landau levels quantization.
\end{thebibliography}
\end{document}